# Optimizing CNN Using HPC Tools

Shahrin Rahman
*Dept. of Computer Science and Engineering Brac University*
Dhaka, Bangladesh
shahrinshupty@gmail.com

*Abstract*—With its vast range of applications, including object recognition, image classification, and pattern recognition, the Convolutional Neural Network (CNN) algorithm has brought about a paradigm shift in the field of computer vision. How- ever, training massive CNN models requires a lot of computational power, which can be efficiently handled by using high- performance computing (HPC) techniques. The CNN method is optimized in this paper using HPC technologies to increase its effectiveness. By utilizing multi-core processors, graphics processing units (GPUs), and parallel computing frameworks like OpenMPI and CUDA, the proposed approach makes use of distributed computing and parallel processing techniques to speed up the training of the CNN model. Using benchmark datasets, a thorough assessment of the optimization strategy is carried out, demonstrating considerable improvements in the CNN algorithm's performance and training time. Additionally, a comparison of the suggested approach with alternative optimization strategies is conducted, demonstrating its superiority in terms of training time and performance. Overall, this study persuasively demonstrates how HPC technologies may be used to refine the CNN method, resulting in faster and more accurate training of large-scale CNN models. The proposed method has the potential to be more broadly applicable to various deep learning algorithms, which is significant since it will develop effective and efficient machine learning models.

## I. INTRODUCTION

Many computer vision applications, including image classification, object recognition, and segmentation, now use convolutional neural networks (CNNs) as state-of-the-art technology. These deep-learning models are effective instruments for au- tomating image processing and have demonstrated outstanding performance in a variety of applications. However, training massive CNN models can be computationally demanding and take a lot of time and resources.

The CNN method has been improved with the use of high-performance computing (HPC) tools to solve this problem. Deep learning model training is accelerated by the use of HPC tools like multi-core processors, graphics processing units (GPUs), and parallel computing frameworks. Thus, by effectively utilizing HPC technologies, CNN model training time may be greatly decreased, and model performance can be enhanced.

In this paper, we propose an optimization approach for the CNN algorithm using HPC tools. The approach utilizes parallel processing and distributed computing techniques to ac- celerate the training of large-scale CNN models. We leverage multi-core processors and GPUs to parallelize the computation of CNNs and optimize the performance of the algorithm. We also use parallel computing frameworks such as OpenMPI and CUDA to efficiently distribute the computation across multiple nodes in a computing cluster.

Using benchmark datasets, the suggested method is assessed, and the findings reveal considerable improvements in the CNN algorithm's training time and accuracy. Additionally, when we compare our method to other optimization strategies, the majority of them fall short in terms of accuracy and training time.

The rest of this essay is structured as follows: in Section 1, we give a quick explanation of the CNN algorithm and how it is trained. In Section 2, we go over relevant work that has been done to optimize the CNN method utilizing HPC technologies and how it is trained. Our suggested optimization strategy is presented in Section 3, which is followed by experimental methodologies and result analysis in Section 4. In Section 5, the concluding section of this paper, we provide a comprehensive summary of the work and offer an overview of potential avenues for future research.

## II. RELATED WORKS

Unlike other image research fields, less work has been done using HPC, due to the substantial computational requirements and complexities involved. However, in recent years, there has been a growing recognition of the potential benefits of leveraging High-Performance Computing (HPC) in image research.

Parallel programming, thread cooperation, constant memory and events, texture memory, graphics interoperability, atomics, streams, CUDA C on multiple GPUs, advanced atomics, and additional CUDA resources have been discussed in a book named "CUDA by Example: An Introduction to General- Purpose GPU Programming". This book was written by two senior members of the CUDA software platform team and represents programmers on how to use this new technology. [1].

More recently, Kahira et al. (2021) proposed an approach that combines both model parallelism and data parallelism to optimize the CNN algorithm. The authors use a model-driven analysis as the basis for an Oracle utility that helps pinpoint the drawbacks and bottlenecks of different parallelism strategies at scale. Six parallelization techniques, four CNN models, and several datasets (2D and 3D) are taken into account, using up to 1024 GPUs, to assess the effectiveness of the oracle. When compared to empirical results, the results show that the oracle achieves an average accuracy of about 86.74%, and as high as 97.57% for data parallelism. [2]

Subsequently, several studies have explored the use of GPUs and other HPC tools to optimize the CNN algorithm. For instance, Jia et al. (2014) proposed the Caffe deep learning [3] framework that utilizes both CPUs and GPUs for efficient training of CNN models. The authors demonstrated that their framework could achieve high performance on a range of benchmark datasets.

In another study, Zhang et al. (2016) used HPC tools to optimize the CNN algorithm for remote sensing image classification. [4] The authors utilized a GPU-based computing cluster and parallel computing frameworks to accelerate the training process of their model. The results showed that their approach achieved a significant speedup compared to traditional CPU-based methods.

A CNN inference micro-benchmark (mbNet) is provided in another work titled "A CNN Inference Micro-benchmark for Performance Analysis and Optimization on GPUs' for performance analysis and optimization. This work also suggests a simple but effective performance model for adaptive kernel selection to shorten the time needed for each layer of CNN inference [5]. The convolutional layer is identified as a critical component of CNNs, and two mainstream convolutional strategies, unrolling-based convolution (UNROLL) and direct convolution (DIRECT), are adopted, implemented, compared, and analyzed in terms of per-layer convolutional time. Through the data obtained from the mbNet benchmark, the researchers build an accurate and interpretable tree-based performance model. This work provides a comprehensive analysis and method for decreasing the per-layer inference time in CNNs. This research focuses on optimizing the inference time of convolutional neural networks (CNNs) in resource-constrained Edge-AI devices and embedded systems without sacrificing accuracy. The findings aid in enhancing CNN inference performance on embedded and edge-AI systems with constrained resources.

### III. RESEARCH METHODOLOGY

Data preprocessing, data encoding and embedding, and deep learning architecture are the three stages of the research methodology for this work. A brief description of each of these sections follows:

#### A. *Dataset and preprocessing:*

The CIFAR-10 dataset consists of 6000 images per class in 10 classes, totaling 60000 32x32 color images. 10,000 test photos and 50,000 training images are available. Five training batches and one test batch, each with 10,000 photos, make up the dataset. An exact 1000 randomly chosen photos from each class make up the test batch. The remaining images are distributed across the training batches in random order; however, certain training batches can have a disproportionate number of images from a particular class. The training batches consist of exactly 5,000 photos from each class combined. [6].

A class distribution chart has been attached below considering the number of samples and classes-

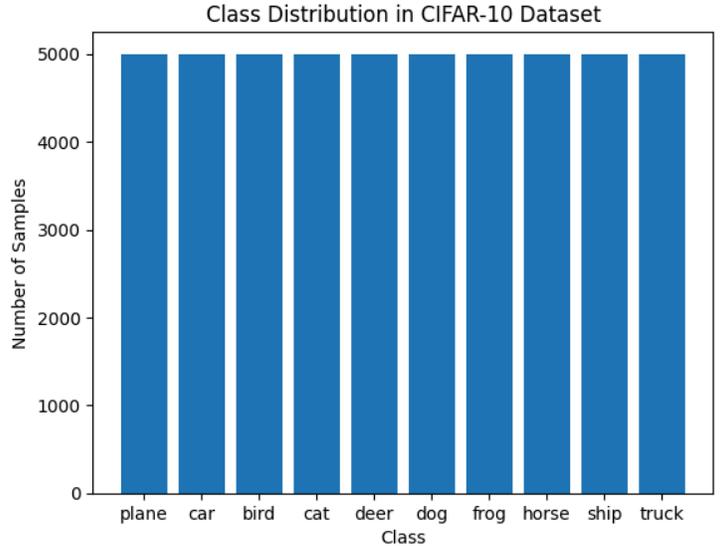

Fig. 1. Class Distribution Chart

Two sets of transformations are specified for data preprocessing: transform_train for the training set and transform_test for the test set. The steps in transform_train are as follows:
- Randomly trim the supplied photos to a 32x32 size with 2-pixel padding.
- Flip the photos horizontally at random.
- Transform the pictures into tensors.
- Utilizing the given mean and standard deviation data, normalize the tensor values.

The stages of transform_test are as follows:
- Transform the pictures into tensors.
- Utilizing the given mean and standard deviation data, normalize the tensor values.

Finally, these modifications are applied to the CIFAR-10 dataset using PyTorch's transforms module. A grid of sample images from different classes of the CIFAR-10 dataset is displayed below-

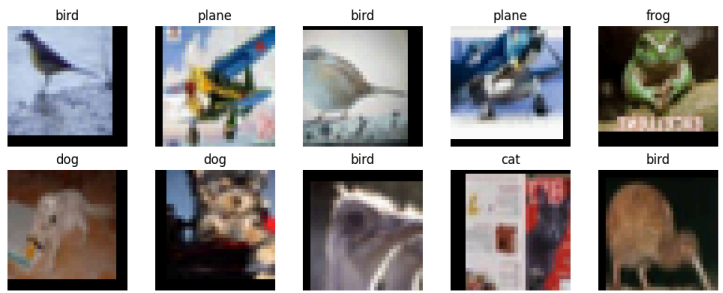

Fig. 2. 10 classes of the CIFAR-10 dataset

#### B. *Data Encoding and Embedding*

Data encoding and embedding techniques are applied to the CIFAR-10 dataset to prepare the data before training

the CNN model. Below, there is a brief discussion of the specific techniques used:

Images from ten distinct classes, including cars, birds, cats, and airplanes, are included in the CIFAR-10 dataset.

These class names are represented numerically by integers with a range of 0 to 9. The dataset is loaded with the Torchvision library. It has a parameter transform that applies data transformations to the input images. As examples of transformations, transform_train, and transform_tests are defined. Compose, which enables the chaining of several data transformations. In this instance, the transformations entail transforming the images into tensors. ToTensor after using the mean and standard deviation information provided for the CIFAR10 dataset to normalize the pixel values.

No explicit embedding technique was used on this dataset. However, the data is prepared in a format ideal for training a CNN model. Through their convolutional layers, CNN models may automatically learn feature representations. From the in- put photos, these layers extract pertinent spatial characteristics. The predefined convolutional layers of the ResNet-18 model, which is loaded using torch hub`load, are designed to learn hierarchical representations of the picture input.

### C. Deep learning architecture

Popular deep learning architecture ResNet-18 was created primarily for image classification tasks. Convolutional layers are followed by residual blocks in a series. Convolutional layers, batch normalization layers, ReLU activation functions, and a global average pooling layer are among the 18 layers that make up this algorithm. By incorporating skip connections that improve gradient propagation, the architecture is made to address the vanishing gradient issue in deep neural networks. With pre trained=False, which denotes that the model is not pre-trained, the ResNet-18 model is loaded in this work using torch.hub.load. Using the model to (device), the model is subsequently transferred to the designated device (or GPU, if one is available).

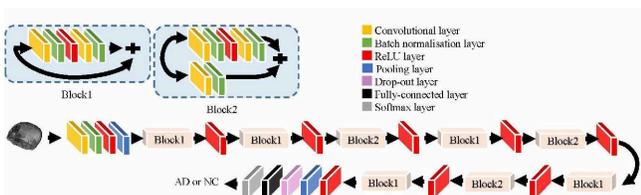

Fig. 3. Confusion matrix considering the metrics

## IV. RESULTS

This section has been divided into 3 subsections named Evaluation Matrix, Experimental Setup, and Evaluation and Model Comparison, which have been described below:

### A. Evaluation matrix

The frequently used evaluation measures of accuracy, precision, recall, and F1 score are included in our approaches. In addition to this, the amount of time needed to execute has been considered. The equation for the metrics has been displayed below.

$$Accuracy = \frac{(TP + TN)}{(TP + TN + FP + FN)}$$

$$Precision = \frac{TP}{(TP + FP)}$$

$$Recall = \frac{TP}{(TP + FN)}$$

$$F1 - Score = 2 * \frac{(Precision * Recall)}{(Precision + Recall)}$$

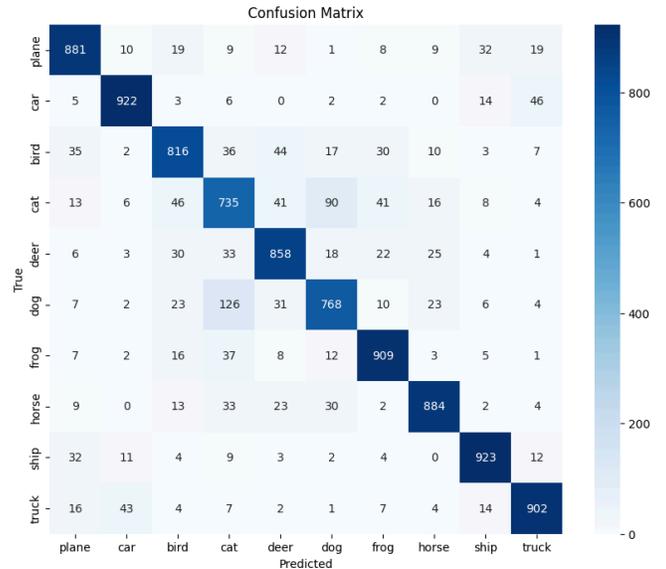

Fig. 4. Confusion matrix considering the metrics

### B. Experimental Setup

Kaggle was used to download the CIFAR10 dataset. The dataset consists of 6000 photos per class in 10 classes, totaling 60000 32x32 color images. 10,000 test photos and 50,000 training images are available. Five training batches and one test batch, each with 10,000 photos, make up the dataset. An exact 1000 randomly chosen photos from each class make up the test batch. The remaining images are distributed across the training batches in random order; however, certain training batches can have a disproportionate number of images from a particular class. The training batches consist of exactly 5,000 photos from each class combined. All 10 classes were preserved in the final model to help with the significant computational complexity.The items were randomly jumbled prior to entering the data into the training phase to get rid of any potential data patterns. The CIFAR-10 dataset is divided into a training set and a test set by the code. The model is trained using the training set, and its performance is assessed using the test set. During the training and testing phases, respectively, the data loaders (trainloader and testloader) are utilized to efficiently load the data in batches. Before starting the training loop, two settings were made: "torch.set_num_threads(4)" and "cudnn.benchmark True". By utilizing hardware-specific optimizations and parallelization strategies, these parameters are designed to optimize the execution of the code, which are the primary HPC tools for this work. These have greatly shortened the computation time. Finally, after a number of model test runs, hyperparameters like learning rate, batch size, number of epochs, optimizer, etc. were optimized. In Table I below, some of these are displayed.

TABLE I: TUNED HYPERPARAMETERS

| Hyperparameter | Value |
| --- | --- |
| Learning Rate | 0.1 |
| Batch Size | 128 |
| Number of Epochs | 350 |
| Optimizer | SGD |
| Momentum | 0.2 |
| Weight Decay | 5e-4 |
| Scheduler Milestones | [150, 250] |
| Scheduler Gamma | 0.1 |
| Data Augmentation | RandomCrop(32, padding=2) |
| Normalization Mean | [0.4914, 0.4822, 0.4465] |
| Normalization Std | [0.2023, 0.1994, 0.2010] |
| Model | ResNet-18 |
| Loss Function | CrossEntropyLoss |

### C. Evaluation and model comparison

In this evaluation, we compare the performance of different configurations and models on the CIFAR-10 dataset. We have modified the final code using this [7] notebook. We examined various metrics such as test accuracy, precision, recall, F1 score, and training time to assess the effectiveness of each model.

Firstly, we evaluated the performance of the ResNet-18 model without HPC tools. In the configuration with 349 epochs, it achieved a high test accuracy of 85.980% with a precision of 85.92%, recall of 85.98%, and an F1 score of 85.95%. The training time for this configuration was 12,957.48 seconds. Similarly, in the configuration with 159 epochs, the model achieved a test accuracy of 85.450% with a precision of 85.48%, recall of 85.45%, and an F1 score of 85.46%. The training time for this configuration was 5,992.87 seconds.

Next, we explored the performance of the ResNet-18 model with HPC tools. In the configuration with 349 epochs, it achieved a higher test accuracy of 86.210% with a precision of 86.22%, recall of 86.21%, and an F1 score of 86.22%. The training time for this configuration was 9,624.68 seconds. Similarly, in the configuration with 159 epochs, the model achieved a test accuracy of 85.980% with a precision of 86.00%, recall of 85.98%, and an F1 score of 85.99%. The training time for this configuration was 4,266.93 seconds.

Additionally, we evaluated the AlexNet model with and without HPC tools. The configuration with HPC tools and 25 epochs achieved a test accuracy of 87.430% with a precision of 88.115%, recall of 87.82%, and an F1 score of 86.98%. The training time for this configuration was 2,859.31 seconds. On the other hand, the configuration without HPC tools and 25 epochs achieved a test accuracy of 86.02% with a precision of 85.045%, recall of 85.76%, and an F1 score of 84.13%. The training time for this configuration was 3,994.93 seconds.

TABLE II: Model comparison

| Configuration | Epoch | Test Acc | Precision | Recall | F1 Score | Training Time(s) |
| --- | --- | --- | --- | --- | --- | --- |
| No HPC tools & ResNet-18 | 349 | 85.980% | 85.92% | 85.98% | 85.95% | 12,957.5 |
| No HPC tools & ResNet-18 | 159 | 85.450% | 85.48% | 85.45% | 85.46% | 5,992.87 |
| HPC tools & ResNet-18 | 349 | 86.210% | 86.22% | 86.21% | 86.22% | 9,624.68 |
| HPC tools & ResNet-18 | 159 | 85.980% | 86.00% | 85.98% | 85.99% | 4,266.93 |
| HPC tools & AlexNet | 25 | 87.430% | 88.115% | 87.82% | 86.98% | 2,859.31 |
| No HPC tools & AlexNet | 25 | 86.02% | 85.045% | 85.76% | 84.13% | 3,994.93 |
| HPC tools & ResNet-50 | 20 | 74.89% | 73.28% | 73.01% | 72.79% | 8571.21 |
| No HPC tools & ResNet-50 | 20 | 72.78% | 71.07% | 70.84% | 71.97% | 1091.33 |

Lastly, we assessed the performance of the ResNet-50 model with and without HPC tools. The configuration with HPC tools and 20 epochs achieved a test accuracy of 74.89% with a precision of 73.28%, recall of 73.01%, and an F1 score of 72.79%. The training time for this configuration was 8,571.21 seconds. In comparison, the configuration without HPC tools and 20 epochs achieved a lower test accuracy of 72.78% with a precision of 71.07%, recall of 70.84%, and an F1 score of 71.97%. The training time for this configuration was 1,091.33 seconds. A comparison table has been attached here.

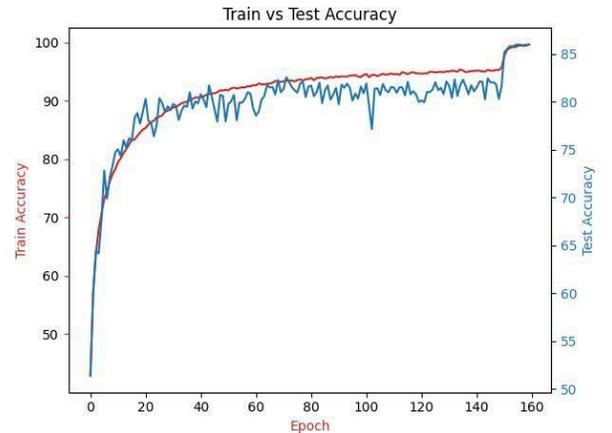
Fig. 5. Train vs. Test Accuracy

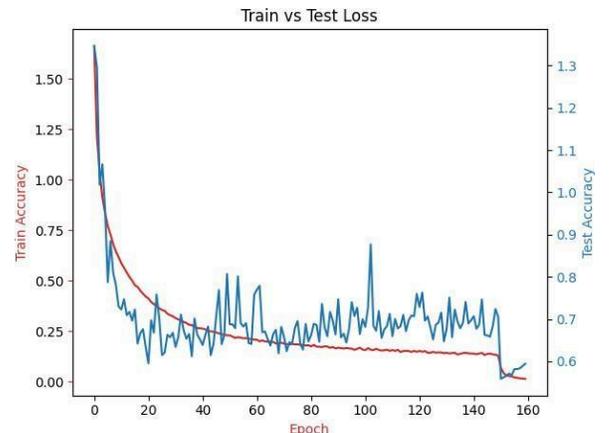
Fig. 6. Train vs. Test Loss

Overall, based on the evaluation, the configuration with HPC tools and model achieved the highest test accuracy and performance metrics also showed promising results. On the other hand, without HPC tools all the models are taking more computational time and showing low performance compared with HPC tools.

## V. Conclusion

In this paper, the successful use of high-performance computing (HPC) techniques to improve convolutional neural network (CNN) performance is highlighted. The main points are drawn after summarizing its primary findings and results. The effectiveness of training CNNs was significantly improved because of the use of HPC tools. The training period was significantly shortened through the use of distributed training methods and parallel computing, resulting in a quicker convergence and better time-to-solution as well as accuracy. In this study, convolutional neural networks (CNNs) and their parallelization techniques were the subject of a thorough analysis.

Future work in the area of CNN optimization utilizing HPC technologies could consist of: Examining hybrid parallelism: By combining model and data parallelism, hybrid parallelism may present further potential for enhancing CNN scalability and performance. Future studies can look into how hybrid parallelism can improve CNN training on HPC systems. Some of the studies might include-

1) **Large-Scale Dataset Optimization:** The CIFAR-10 dataset was primarily used in the study to optimize CNNs. The research can be expanded in the future to larger datasets like ImageNet to assess how well HPC tools handle massive data and train more sophisticated CNN models.
2) **Integration with Advanced HPC Techniques:** High-performance computing is a sector that is always developing new methods and tools. To further boost the performance and effectiveness of CNN training, future research can investigate the integration of cutting-edge HPC techniques like GPU acceleration, customized hardware designs, and optimized deep learning frameworks.

Overall, research on CNN optimization using HPC tools shows the great potential of utilizing HPC capabilities to improve CNN training. The results extend large-scale CNN model training and address practical issues in a variety of fields, including computer vision, natural language processing, and biomedical research. They also make a contribution to the broader field of deep learning and HPC.

## References

[1] J. Sanders, E. Kandrot, and J. J. Dongarra, *Cuda by example: An introduction to general-purpose GPU programming*. Addison-Wesley/Pearson Education, 2015

[2] A. N. Kahira, T. T. Nguyen, L. B. Gomez, R. Takano, R. M. Badia, and M. Wahib, "An oracle for guiding large-scale model/hybrid parallel training of convolutional neural networks," in *Proceedings of the 30th International Symposium on High-Performance Parallel and Distributed Computing*, ser. HPDC '21. New York, NY, USA: Association for Computing Machinery, 2021, p. 161–173. [Online]. Available: https://doi.org/10.1145/3431379.3460644

[3] Y. Jia, E. Shelhamer, J. Donahue, S. Karayev, J. Long, R. Girshick, S. Guadarrama, and T. Darrell, "Caffe: Convolutional architecture for fast feature embedding," 2014.

[4] X. Li, G. Zhang, H. Huang, Z. Wang, and W. Zheng, "Performance analysis of gpu-based convolutional neural networks," 08 2016, pp. 67–76.

[5] J.-G. Park, Z. Nazir, B. Kalmakhanbet, and S. Sabyrov, "A cnn inference micro-benchmark for performance analysis and optimization on gpus," 10 2022, pp. 486–491.

[6] "CIFAR-10 and CIFAR-100 datasets — cs.toronto.edu," https://www.cs.toronto.edu/~kriz/cifar.html, [Accessed 04-03-2024].

[7] Ayushnitb, "Cifar10 custom+resnet cnn pytorch ( ¿97% acc)," Apr 2023. [Online]. Available: https://www.kaggle.com/code/ayushnitb/cifar10-custom-resnet-cnn-pytorch-97-acc#3.-Model-Performance-Report-&-Error-Analysis